\documentclass[10pt,conference]{IEEEtran}

\usepackage{graphics} % for pdf, bitmapped graphics files
\usepackage[pdftex]{graphicx}
\usepackage{times} % assumes new font selection scheme installed
\usepackage [cmex10]{amsmath} % assumes amsmath package installed
\usepackage{amssymb}  % assumes amsmath package installed
\usepackage[compress]{cite}
 \IEEEoverridecommandlockouts
\title{Capacity of Gaussian MAC Powered by Energy Harvesters without Storage Buffer
\thanks{ R Rajesh is  with Center for Airborne Systems, DRDO, Bangalore. Deekshith P K and Vinod Sharma are with ECE dept. Indian Institute of Science, Bangalore. Email:rajesh81r@gmail.com,  deeks@ece.iisc.ernet.in, vinod@ece.iisc.ernet.in}
\author{R Rajesh, Deekshith P K and Vinod Sharma}
\thanks{This work is partially supported by a grant from ANRC to Prof. Sharma.}
}
\begin{document}
\maketitle
\thispagestyle{empty}
\pagestyle{empty}
\begin{abstract}
We consider a Gaussian multiple access channel (GMAC) where the users are 
sensor nodes powered by energy harvesters. The energy harvester has no buffer to store the harvested energy and hence the energy need to be expended immediately. We assume that the decoder has perfect knowledge of the energy harvesting process.  We characterize the capacity region of such a GMAC. We also provide the capacity region when one of the users has infinite buffer to store the energy harvested. Next we  find the achievable rates when the energy harvesting information is not available at the decoder.
\end{abstract}
\noindent
\textbf{Keywords:}  Energy harvesting, Gaussian multiple access channel, Shannon capacity.
%\end{keywords}

\section{Introduction}
\label{intro} 

Wireless sensors equipped with energy harvesting mechanisms are gaining popularity as they  improve the network life time and support 'green communication' \cite{ch-ver-hed-col-pic-dem-14,ch-oh-krish-lin-ni-I,ch-kan-hsu-jah-sri-8,ch-rag-gan-sri-10}. Such wireless sensor networks are used commonly for  random event detection and are designed to work without battery change for many years. The sensors usually transmit on occurrence of the random event and in such cases the energy harvester powering the sensor does not usually store energy. This is because now one can use supercapacitors to store energy instead of rechargeable batteries \cite{bader}.  This makes the design of harvesting mechanism simpler and reduces the cost. Furthermore, sensors may be organized in a hierarchical fashion for such an event detection goal. Multiple access channels are the usual building blocks for such a channel (\cite{af}, \cite{baek}).% ([6], [7])

In this paper we study a Gaussian Multiple Access Channel (GMAC) formed by energy harvesting sensor nodes without  energy buffers to store the harvested energy. Thus the transmission from these nodes  is inherently amplitude limited at each time instant.  We derive the capacity of such a GMAC when the energy harvesting information is available at the decoder also. We also characterize the capacity of the GMAC in the heterogeneous set up when the energy harvester powering one node has `no-buffer' and the other node has `infinite buffer'. Furthermore we provide achievable rates when we relax the assumption that the information about the energy harvesting process is available at the decoder.
%assumption about energy harvesting process being available at the decoder is relaxed.

We survey the related literature. Information capacity of a Gaussian channel with an energy harvesting sensor node is provided in \cite{uluk} and \cite{rajesh}. \cite{rajesh} also provides capacity for a Gaussian channel with an energy harvester without a storage buffer and when a significant amount of energy is consumed in data processing and sensing and when there are  inefficiencies in energy storage. The results in \cite{rajesh} are extended to fading channels in \cite{ch-ra-sh-vi-11A} and when there is bursty traffic to the data queue in \cite{asil}. %\textbf{\cite{ulukasyl} consider an AWGN channel with time varying amplitude constraint as an abstraction of a channel with energy harvesting transmitting nodes with no battery for energy storage.}

The capacity of an AWGN channel with a peak power constraint is provided in \cite{peak1}.  It was shown that the capacity achieving input distribution is discrete with a finite support. Furthermore,  \cite{peak2} considered an amplitude and variance constrained quadrature Gaussian channel and provided the capacity achieving  input distribution. Reference \cite{achan} provides the class of noise densities for which the capacity achieving input distribution is discrete. 

The capacity region  of a  MAC and  of a  GMAC are provided in \cite{cover}. The capacity of a  MAC with peak power constraints   and an infinite number of users is provided in \cite{verdu}. In \cite{ita12} we obtain the results for a GMAC with energy harvesting users which have infinite buffers or finite (non zero) buffers. These results will be summarized in this paper for completeness.  
%Necessary and  sufficient conditions for discreteness of capacity achieving distribution is also provided in \cite{fei}.
%The capacity with partial  channel state information at both the encoder and the decoder for a p%oint to point channel is provided in 

The capacity of a state dependent point to point channel when partial channel state information is available at both the encoders and the decoder is provided in  \cite{caire}. Reference \cite{lapidoth1} provide the achievable region of a MAC when the channel state information is available only at the encoders.

The rest of the paper is organized as follows. In Section II we present our model and also present the capacity for a GMAC with infinite buffers. Section III studies GMAC formed by sensor nodes without energy buffers and establishes the discreteness of the capacity achieving input distributions. It is assumed that the energy harvesting information is available at  both the encoders and the decoder. Section IV obtains the capacity of a GMAC with one node without an energy buffer and the other node with an infinite buffer.  Section V studies the case when the energy harvesting information is available partially at the encoders and the decoder and provide achievable rates. Energy harvesting information available only at the encoders and not at the decoder follows as a special case. Section VI concludes the paper.
 
\section{ Model and notation } 
\label{Section3}
%\subsection{Model and notation}
%\label{model2} In this section we present our model for energy harvesting sensor nodes without storage buffer and transmitting over a GMAC.
\begin{figure}[h]
\begin{center}
\includegraphics[height=1.6in, width=3.2in]{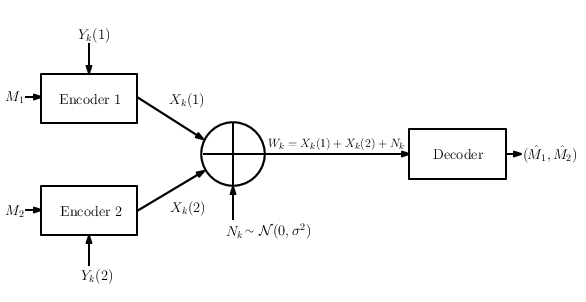}
\caption{The model} \label{fig1}
\end{center}
\end{figure}%(\cite{ch-kan-hsu-jah-sri-8},~\cite{ch-rag-gan-sri-10})
Initially we consider two energy harvesting sensor nodes which are sensing and generating data to be transmitted to a central node via a discrete time GMAC. We assume that transmission consumes most of the energy in the sensor node and ignore other causes of energy consumption like sensing, processing and receiving from other nodes. This is often the case in sensor motes ([3], [4]).  The sensor nodes are indexed by $i=1,2$. At any time instant $k \geq 1$, let  $Y_{k}(i)$  denote the energy harvested by  node $i$. Since there is no storage of energy, the energy has to be used immediately for transmission. The node uses $T_k(i)$ energy at time $k$. Hence we have $T_k(i) \leq Y_k(i),~ k \geq 1$. Let $X_k(i),~ k \geq 1$, denote the transmitted codeword from node $i$. We have $T_k(i)= X_k^2(i)$.

We consider $\{Y_k(i),~{k\geq1}\},$ to be $i.i.d$. The channel output at $k^{th}$ instant is $W_k$ and is equal to $X_{k}(1)+X_{k}(2)+N_k$ where $\{N_k\}$ is independent, identically distributed ($i.i.d$) with mean zero and variance $\sigma^2$ (The corresponding Gaussian density is denoted by $\mathcal{N}(0,\sigma^2))$.

We denote the availability of $Y_k(i)$ information at the encoders and the decoder by CSIT and CSIR respectively. We will say that the rates $(R_1,R_2)$ are achievable for the above MAC if there exist $(2^{nR_1},2^{nR_2},n)$ encoders and a decoder such that the average probability of error $P_{e}^{(n)} \rightarrow 0$ as $n \rightarrow \infty$. In \cite{ita12} we have obtained the capacity region of this system when all the nodes have infinite buffer length. There it is shown that the capacity region of the system is the same as of a usual GMAC where the users are connected to a regular power supply where user $i$ has average power constraint $\mathbf{E}[Y(i)]$. In \cite{ita12} achievable rate region for the GMAC are also obtained when the energy buffer at each node are finite. In this paper we consider its special case when there are no energy storage at any node. In this special case we can obtain the capacity region of the GMAC. We will also consider the case when some of the nodes have infinite buffer and some do not have any.

Although, for simplicity, we present results for two users, all results in this paper extend to multiuser case without any difficulty. 

\section{Sensor nodes without energy buffer}

\subsection{GMAC with peak power constraints}
When the sensor nodes do not have energy buffers, the nodes become  peak power constrained with the constraint varying from slot to slot according to $\{Y_k(i),{k \geq 1}\}$. Hence, initially, we study a two user GMAC with peak power constraints. Let  the   peak power constraint on node $i$ be $P(i)$.  Let $A(i)$ denote the equivalent amplitude constraint, i.e., $A(i)=\sqrt{P(i)}$. 

We are interested in characterizing  the capacity achieving input distributions that achieve the boundary points of the capacity region (Fig. 2). If the noise variance is 1 and input signal amplitude is low, typically, less than 1.68  (\cite{nareshncc}), binary signalling with two mass points at the extreme values achieves the boundary points \cite{verdu}. We investigate the 2-user GMAC with arbitrary peak power constraints. We also extend the result to the M-user case.

\emph{\textbf{Theorem 1:}} For a 2-user GMAC with the $i^{th}$ user peak amplitude constrained by $A (i)$, $i=1,2$, boundary points of the capacity region are achieved by discrete random variables.
\textsl{Proof:} For a 2-user MAC, we have the capacity region ([15])% \cite{cover}$(R_1,R_2)$ given by
\begin{align*}
R_1 < I(X(1);W|X(2)),\\
%=h(X(1)+N)-h(N)\\
R_2 < I(X(2);W|X(1)),\\
%=h(X(2)+N)-h(N)\\
R_1+R_2 < I(X(1),X(2);W),
%=h(X(1)+X(2)+N)-h(N)
\end{align*}
where $X(1)$ and $X(2)$ are independent random variables.

The point D in Fig. 2 is obtained by taking $X(1)$ to be  discrete with the distribution that achieves capacity for an AWGN channel with noise variance $\sigma^2$ and peak power constraint $P(1)$. Similarly we obtain point A.

The boundary points B and C in Fig. 2 are obtained via successive cancellation and any point on B-C segment is obtained via time sharing.

\begin{figure}[h]
\begin{center}
%\hspace{50pt}
\includegraphics[scale=.4]{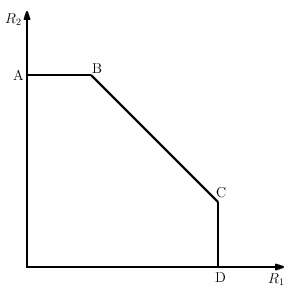}\label{fig_one}
\caption{MAC capacity region}
\end{center}
\end{figure}

%\cite{nareshncc}
Now we consider the input distributions to achieve point B  for node 1 when $X(2)$ is transmitting with a discrete distribution with a finite number of mass points in $[-A(2),+A(2)]$ to achieve A. $X(2)+N$ is the noise for $X(1)$ at B and $X(i),~ i=1,2$ is independent of $N$. Since $X(2)$ is  finite and symmetric about the origin ([20]) and $N$ is Gaussian,  $X(2)+N$ has a density denoted by $f_{X(2)+N}(.)$ which is a mixture of Gaussian densities with  mean values symmetric about the origin. Using the results from \cite{achan} we show that user 1 needs a discrete distribution.

We show that $f_{X(2)+N}(.)$ belongs to the noise class described in \cite{achan}. The density function of a Gaussian mixture noise admits an analytic extension over the entire complex plane, given by
\begin{eqnarray*}
f_{X(2)+N}(z)=\sum_{i=1}^{n}\frac{p_i}{\sqrt{2\pi\sigma^2}}e^{-\frac{(z-\mu_i)^2}{2\sigma^2}}
\end{eqnarray*}
The four conditions given in \cite{achan} are satisfied by defining functions
\begin{eqnarray}
U(x) &=&\frac{1}{\sqrt{2\pi\sigma^2}}e^{\frac{-x^2+2x\mu_{max}}{2\sigma^2}} ,~ x \geq k >2\mu_{max},\\
L(x)&=&  \frac{\epsilon}{\sqrt{2\pi\sigma^2}}e^{\frac{-x^2-\mu_{max}^2+2x\mu_{min}}{2\sigma^2}} ,~ x \geq k,
\end{eqnarray}
where $\mu_{max}$ and $\mu_{min}$ are the maximum and minimum values of the mean values of  the component Gaussian densities in the mixture. Hence, as proved in \cite{achan},   point  B  is achieved by a discrete  signalling with a finite support. 

By symmetry, the same conclusion holds for point C. It may be noted that, the distribution of node $1$ (or 2) achieving point B (or C) will be different from the one that achieves D (or A). $~~~~~~~~~~~~~~~~~~~~~~~~~~~~~~~~~~~~~~~~~~~~~~~~~~~~~~~~~~~~~~~~~~~~\blacksquare$% The non-corner points in the boundary on line BC be can be achieved by time sharing. 

For more than two users, the capacity region of a MAC corresponding to Fig. 2 is available in \cite{cover}. The boundary points of it are obtained by successive decoding and time sharing. Thus, as in proof of Theorem 1 above we can show that for this case also the boundary points for a GMAC with peak power constraints are obtained by discrete distributions.

\subsection{GMAC with energy harvesting sensor nodes}
From Theorem 1 we immediately obtain the following theorem.

\emph{\textbf{Theorem 2:}} The capacity region of a two user Gaussian multiple-access channel powered by  energy harvesting  nodes without a storage buffer for energy storage, and the instantaneous energy harvesting information of both the nodes  known at the encoders and  the decoder, is the closure of the convex hull of all $(R_1,R_2)$ satisfying\\
\begin{equation}
R_1 < \mathbf{E}_{Y(1),Y(2)}[I\big(X_1{(Y_1)};W|X_2(Y_2)\big)],
\end{equation}
\begin{equation}
R_2 < \mathbf{E}_{Y(1),Y(2)}[I\big(X_2(Y_2);W|X_1(Y_1)\big)],
\end{equation}
\begin{equation}
R_1+R_2 < \mathbf{E}_{Y(1),Y(2)}[I\big(X_1(Y_1),X_2(Y_2);W\big)],
\end{equation}
{where we denote the random variable $X(i)$  satisfying the peak power constraint $Y(i)$ compactly by $X_i(Y_i)$ and $X_1(Y_1)$ is independent of $X_2(Y_2)$}. The boundary points are achieved by discrete independent input random variables $X(1)$ and $X(2)$.$~~~~~~~~~~~~~~~~~~~~~~~~~~~~~~~~~~~~~~~~~~~~~~~~~~~~~~~~~~~~~~~ \blacksquare$

Corresponding results holds for more than two users also.
% are independent r.v.s satisfying the peak power constraints $Y(1)$ and $Y(2)$.\\
%\textsl{Proof:} For simplicity, let the state space $\mathcal{Y}(1), \mathcal{Y}(2)$ be finite. This result can be easily extended to the continuous case.  Let $ y_{k}(1), ~k =1,..., N \in  \mathcal{Y}(1)$ and  $y_{k}(2), ~k =1,..., M \in  \mathcal{Y}(2)$.  We model each realization of $Y_k(i)$ as a channel side information. Thus the channel is a state dependent GMAC.  Since the side information $Y_k(i), ~i=1,2$ is universally known, Theorem 1 gives that boundary points of the capacity region for each $Y_k(i)$ and  is achieved by discrete random variables at any time instant. Averaging over the  $Y_k(i),~i=1,2$ process gives the desired result.

\section{GMAC with no buffer -infinite buffer combination}
\label{Section4}
%\begin{figure}[h]
%\begin{center}
%\includegraphics[height=2in, width=3in]{ita3nn}
%%\includegraphics[scale=0.45]{single_node.eps}
%\caption{The model} \label{fig2}
%\end{center}
%\end{figure}

In this section we consider a GMAC with node 1 without buffer  and node 2 with infinite buffer to  store the harvested energy. The sensor node 2 is able to replenish energy by $Y_k(2)$  at time $k$. The energy available in the node at time $k$ is $E_k(2)$.  %We will consider the effect of finite capacity later.

Node 2 uses energy $T_k(2)$ at time $k$ which depends on $E_k(2)$ and $T_k(2) \le E_k(2)$ . The process $ \{ E_k(2) \}$ satisfies
\begin{eqnarray}
E_{k+1}(2)  = (E_k(2) - T_k(2))^{+} + Y_k(2). \label{eqn2}
\end{eqnarray}
Rest of the notation is as in Section II.

\emph{\textbf{Theorem 3:}} The capacity of a two user GMAC with  node 1 without buffer and node 2 with infinite buffer and the instantaneous energy harvesting of user 1 known to encoder 2 and to the decoder, is the closure of the  convex hull of all $(R_1,R_2)$ satisfying\\
\begin{eqnarray}
R_1 &<& \mathbf{E}_{Y(1)}[I\big(X_1(Y_1);W|X(2)\big)],\label{one}\\
R_2 &<& \mathbf{E}_{Y(1)}[I\big(X(2);W|X_1(Y_1)\big)],\label{two}\\
R_1+R_2 &<& \mathbf{E}_{Y(1)}[I\big(X_1(Y_1),X(2);W\big)],\label{three}
\end{eqnarray}
where {$X(1)$ with the peak power constraint $Y(1)$ is denoted as $X_1(Y_1)$ and $X(2)$ is a random variable with average power constraint $\mathbf E[Y(2)]$. Also, $X_1(Y_1)$ and $X(2)$ are independent of each other.}$~~~~~~~~~~~~~~~~~~~~~~~~~~~~~~~~~~~~~~~~~~~~~~~~~~~~~~~~~~~~~~~~~~\blacksquare$

%\textit{Proof:}
%\newline the\eqref{one}
At any time instant, user 1 is limited by a peak amplitude constraint. As per Fig. 2 point D corresponds to $I\big(X_1(y_1);W|X(2)\big)$ for any realization of $Y_k(1)=y_1$ and is achieved by the discrete point to point capacity achieving  signalling with peak power $y_1$.  User 2 has an infinite energy buffer. Hence point A corresponding to \eqref{two} is achieved by using the signalling scheme as  in \cite{rajesh}:  $X_k(2)= sgn (X'_k(2)) \min( \sqrt{E_k(2)},|X'_k(2)|)$ where $sgn(x)=1$ if $x\ge0$ and $=-1$ if $x<0$.  Also, $\{X_k',{k \geq 1}\}$ is an $i.i.d$
Gaussian sequence with mean zero and variance $E[Y(2)]-\epsilon$ where
$\epsilon > 0$ is an arbitrarily small constant.
% Then $T_k(2)=X_k(2)^2 \le E_k(2)$ and $E[T_k(2)]=E[X_k(2)^2]\le E[Y(2)]-\epsilon$. Thus $E_k(2) \to
%\infty~a.s.$ and $|X_k(2)-X_k(2)'| \to 0~a.s$. Hence the hard energy constraints of user 2 do not  affect the capacity achieving distribution of point A.   

Point B is achieved by user 2 using the same (truncated) Gaussian signalling used to achieve point A and user 1 using the optimum discrete distribution corresponding to peak power constraint $Y(1)$ and an AWGN channel with variance $\sigma^2+E[Y(2)]$.  

The important case is pertaining to point C. User 1 is transmitting at the optimum discrete distribution to obtain point D. Thus the noise term corresponding to user 2 is $X_1+N$ which is a mixture of Gaussian random variables with different means and the same variance. Therefore, Gaussian codebook for user 2 is not  capacity achieving. In fact we show in Appendix A that the capacity achieving distribution is discrete.

 Non-corner points can be achieved by time sharing and decoding is via joint typicality (or successive cancellation) as CSIR is available. 

% The capacity of a two user Gaussian multiple-access channel where one user is  powered by an energy harvesting  node without buffer and the other user is powered by a constant power supply, with an average power constraint P(2) and the instantaneous energy replenishment of user 1 known to the encoder 2 and to the decoder, is the convex hull of all $(R_1,R_2)$ satisfying \eqref{one}-\eqref{three} with $\mathbf{E}[Y(2)]$ replaced by $P(2)$.\\
 
 \emph{Example 1:}
 \begin{figure}[h]
\begin{center}
\includegraphics[height=1.7in, width=3in]{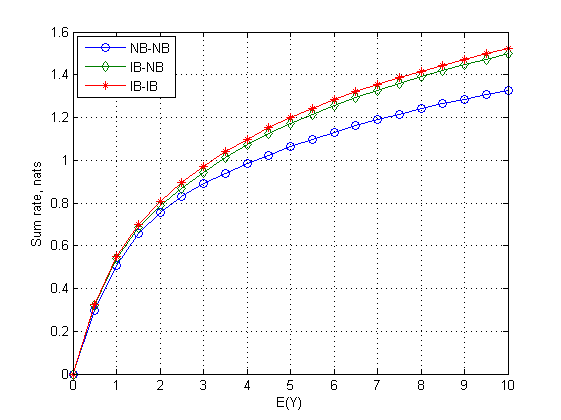}
\caption{Comparison of sum rate of a GMAC with IB-IB,IB-NB,NB-NB} \label{fig5}
\end{center}
\end{figure}
We consider the case where $\sigma^2=1, Y_k(i)=Y_k,~ i=1,2$. $Y_k$ takes values in the alphabet $\mathcal{Y}=\{y_1,y_2\}$ with uniform probability. We vary $y_1,y_2$ to obtain different $\mathbf{E}[Y]$ values.  We plot in Fig. 3 the maximum sum rate when both users have no buffer (NB-NB), when both users have infinite buffer (IB-IB) and when one has infinite buffer and the other has no buffer (IB-NB).
%\begin{eqnarray*}
%R_1 &<& \mathbf{E}_{Y(1)}[I\big(X(Y_1);W|X(2)\big)]\\
%R_2 &<& \frac{1}{2}\log\left(1+\frac{P(2)}{\sigma^2}\right)\\
%R_1+R_2 &<& \mathbf{E}_{Y(1)}[I\big(X(Y_1),X(2);W\big)]
%\end{eqnarray*}
%
%In this case the hard energy constraints in Theorem 3 are removed. The $(R_1,R_2)$ region in Theorem 3 does not depend on the hard constraints and the rate region remains same. 

In Fig. 4 for the NB-NB case we compare the sum rate when each node uses the AWGN channel peak power constrained code with the input that optimizes the sum rate. We fix $Y_k(1)=Y_k(2)= \mathbf{E}[Y(i)],~i=1,2$. From the figure we see that optimal point to point AWGN code is suboptimal in the GMAC case. %It shows that using the point to point AWGN- optimal codes is suboptimal in the GMAC case.
\begin{figure}[h]
\begin{center}
\includegraphics[height=1.7in, width=3in]{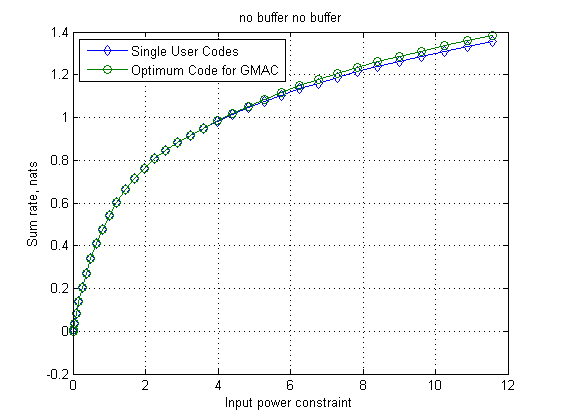}
\caption{Comparison of sum rate of a GMAC with NB-NB} \label{fig5}
\end{center}
\end{figure}
\section{Partial CSIT and CSIR} % {\bf{This section requires a re-look}}
\label{Section5}

In this section, we consider the case when both the users have no buffer. We make our model more realistic by assuming that the energy harvesting information ${Y_{k}(1)}, {Y_{k}(2)}$ is not exactly available at the encoder and the decoder. The model is shown in Fig. 5. The state $(Y_k(1),Y_k(1))$ is available as side information $V_k^{(t)}(i)$ at encoder $i,~i=1,2$. Similarly, at the decoder, another noisy version $\big(V_k^{(r)}(1),V_k^{(r)}(2)\big)$ of the state is available as side information. %We extend the Shannon strategies in \cite{sha1} to find an achievable rate for this channel.

\begin{figure}[h]
\begin{center}
\includegraphics[height=1.5in, width=3in]{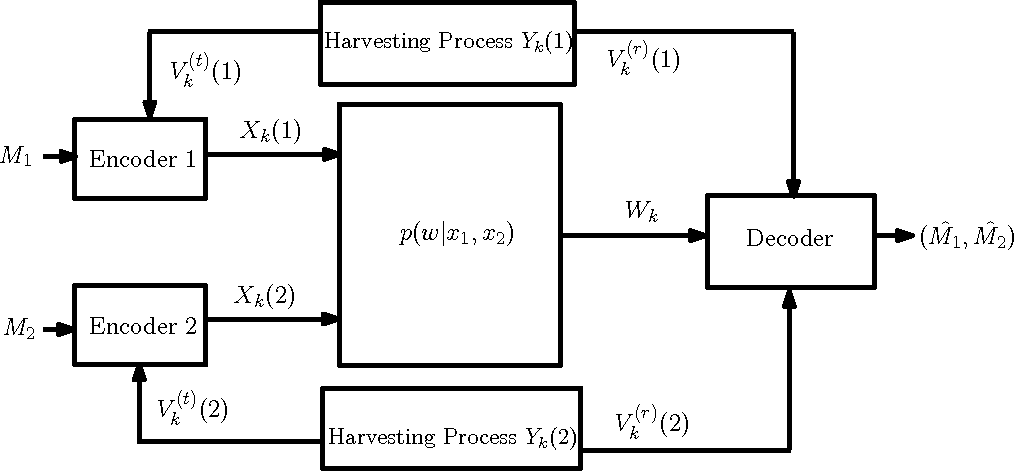}
\caption{The model} \label{fig6}
\end{center}
\end{figure}

An equivalent channel model  for  Fig. 5 is as follows. We assume that a user knows its own energy state perfectly. $\big(V_k^{(r)}(1),V_k^{(r)}(2), W_k\big)$ is taken as the channel output. Then Shannon strategies \cite{sha1} are applied at the each encoder.

Take $U(1),U(2)$ as auxiliary random variables. For $i=1,2$, $g_i: \mathcal{V}^{(t)}(i) \times \mathcal{U}(i) \rightarrow \mathcal{X}(i) $, where $\mathcal{V}^{(t)}(i),\mathcal{V}^{(r)}(i),\mathcal{U}(i),\mathcal{X}(i)$ are the alphabets of $V^{(t)}(i),V^{(r)}(i),U(i),X(i)$ respectively. This equivalent channel model is state independent with  input alphabet $\mathcal{U}(1) \times \mathcal{U}(2)$ and output alphabet $\mathcal{W} \times \mathcal{V}^{(r)}(1) \times \mathcal{V}^{(r)}(2)$. As in\cite{caire}, we have the conditional probability distribution of the output $\big(W,V^{(r)}(1),V^{(r)}(2)\big)$ as 
\begin{equation*}
\tilde{P}_{W,V^{(r)}|X(1),X(2),V^{(t)}} = \sum_{y}p(y){p}_{} (w,v^{(r)}|x_1,x_2,v^{(t)},y)
%W,V^{(r)}|X(1),X(2),V^{(t)},Y(1),Y(2)
\end{equation*}
%P_{Y(1),Y(2)}
where $V^{(r)}=\big(V^{(r)}(1)$ $,V^{(r)}(2)\big),V^{(t)}=\big(V^{(t)}(1),V^{(t)}(2)\big),$ $y= (y_1,y_2)$. 
%\begin{figure}[h]
%\begin{center}
%\includegraphics[height=1.5in, width=3in]{shn_mac3}
%%\includegraphics[scale=0.45]{single_node.eps}
%\caption{Equivalent channel model} \label{fig6}
%\end{center}
%\end{figure}

 % For simplicity we assume the  symmetric case of $Y(i)\equiv Y,i=1,2$. Denote the common state space by $\mathcal{Y}$. 
 
%We have the following achievable region.\\

%\textbf{Theorem 5:} A rate pair $(R_1,R_2)$ is achievable for a state dependent discrete memoryless MAC \big( $\mathcal{X}_1 \times \mathcal{X}_2 \times \mathcal{Y} ,p(w|x_1,x_2,y),\mathcal{W}$\big) with transmitter having access only to noisy state informations $V^{(t)}(1)$ and $V^{(t)}(2)$ , receiver having access to another noisy version $\big(V^{(r)}(1),V^{(r)}(2)\big)$ of the state, such that
Thus, the convex hull of all $(R_1,R_2)$ such that
\begin{equation}
R_1 <I\big(U(1);W|U(2), V^{(r)}\big),\label{aaa}
\end{equation}
\begin{equation}
R_2 < I\big(U(2);W|U(1),V^{(r)}\big),
\end{equation}
\begin{equation}
R_1+R_2 < I\big(U(1),U(2);W|, V^{(r)}\big),\label{ccc}
\end{equation}
for some $P_{U(1),U(2)}(u_1,u_2)=P_{U(1)}(u_1) P_{U(2)}(u_2)$ and functions $g_i,~ i=1,2$ is an achievable region.

We can specialize this result to the case when there is perfect CSIT and no CSIR by taking $V^{(r)} = \phi, V^{(t)}=(Y(1),Y(2) )$ in the achievable region \eqref{aaa}-\eqref{ccc}.

This result can be easily extended to channels with alphabets $\mathcal{X}(i),\mathcal{W},\mathcal{Y}(i)$ the real line \cite{now} and the transmitter subjected to a peak power constraint $P(i),~i=1,2$.

The achievable region remains the same, except for the additional peak power constraint on the channel input symbols. Further, the nature of the optimum distributions achieving the corner points of the capacity region in Fig. 2, is intact with the result given in Theorem 3. 

In some cases, using block Markov coding to convey  compressed state information to the receiver, can provide some what larger rate region than just using Shannon strategy \cite{lapidoth1} under high SNR. In general we conjecture  that achievable region is the union of the achievable regions obtained by the Shannon strategies and the Markov coding.
 \newline
 
  \emph{Example 2:}
  
 Consider a GMAC with two sensor nodes and a common energy harvesting process $\{Y_k\}$ taking values in the state space $\mathcal{Y} =\{0,E\}$ such that $P_Y(E)=\frac{1}{2}$. The corresponding sum rates for a GMAC with IB-IB, NB-NB with full CSIT and CSIR, NB-NB with full CSIT and No CSIR are shown in Fig. 6.
 \begin{figure}[h]
\begin{center}
\includegraphics[height=1.7in, width=3.1in]{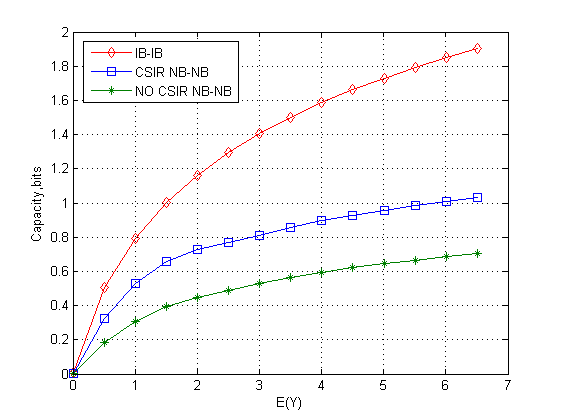}
\caption{ Comparison of sum rates of a GMAC with IB-IB,NB-NB (CSITR),NB-NB(CSIT)} \label{fig5}
\end{center}
\end{figure}
\section{Conclusions}
\label{conclude}
In this paper we consider a GMAC with energy harvesting sensor nodes without an energy buffer. We derive the capacity region assuming perfect causal knowledge of the energy harvesting process at the encoders and the decoder. We also provide the capacity region in the heterogeneous GMAC formed by one node without buffer and another node with infinite buffer. We also extend the results to the case when the information about the energy available at a node is not exactly provided at the encoders and the decoder and provide achievable rate regions.
\section*{Appendix A}
\section*{Proof of the Discreteness of Optimum Distribution given by Theorem 3}

We show that, the optimum distribution for user 2 corresponding to the point C in Fig. 2, is discrete with \textit{at most} a countable number of mass points. For this we show that the capacity achieving input distribution for a scalar additive channel with input second moment constraint P and the noise being a mixture of Gaussian random variables with different means and same variance, is discrete. Let $X$ be the input for this scalar channel, $Y$ be the output and \\
\begin{eqnarray*}
 Y=X+N , \mathbf{E}(X^2) \leq P,
\end{eqnarray*}
where $N$ is $i.i.d$ sequence with distribution
\begin{eqnarray*}
f_N(x)= \sum_{i=1}^{n}\frac{p_i}{\sqrt{2 \pi \sigma^2}}e^{-\frac{(x-\mu_i)^2}{2\sigma^2}}
\end{eqnarray*}
 Let $\mathcal{F}_P \triangleq \{ F: \int x^2 dF(x) \leq P\}$ be the set all distribution functions with second moment constraint P. The capacity of this channel is given by
\begin{eqnarray}
C = \sup_{F \in \mathcal{F}_P} \int\int p(y|x) \log \frac{p(y|x)}{p(y;F)}dy dF(x)
\end{eqnarray} 
\textnormal{where}
\begin{eqnarray}
p(y;F) =  \int p(y|x) dF(x).
\end{eqnarray}   
First, it can be seen that $\mathcal{F}_P$ is convex and compact in the topology of weak convergence. Further, for the above channel model, the mutual information term is continuous in the weak topology, weakly differentiable and strictly concave. Thus it follows from \cite{abou} that a necessary and sufficient condition for the optimum  input distribution $F^*$ is that exists a $\gamma \geq 0$ such that 
\begin{eqnarray}
\gamma(x^2-P) + C -\int p(y|x)\log \big(\frac{p(y|x)}{p(y;F^*)}\big )dy \geq 0
\end{eqnarray}    
for all $x$, with equality if $x$ is in the support of $F^*$.

Let $q(z)=\gamma(z^2-P) + C -\int p(y|z)\log \big(\frac{p(y|z)}{p(y;F^*)}\big )dy$, with 
\begin{eqnarray*}
p(y|z)= \sum_{i=1}^{n}\frac{p_i}{\sqrt{2 \pi \sigma^2}}e^{-\frac{(y-z-\mu_i)^2}{2\sigma^2}} .
\end{eqnarray*} $q(z)$ satisfies the Cauchy-Riemann conditions (For $z=x+iy,~q(z)=u+iv,~$ $\frac{\partial u}{\partial x}=\frac{\partial v}{\partial y}$, $\frac{\partial u}{\partial y}=-\frac{\partial v}{\partial x}$) over the entire complex plane and is analytic everywhere. Let $Z(q)$ denote the set of zeros of $q(z)$. From, \cite{rudin} (Theorem 10.18),  $Z(q)=\mathbb{C}$, the entire complex plane or  $Z(q)$ has no limit point in $\mathbb{C}$ and $Z(q)$ is at most countable. 

Assume that $q(z)= 0$ for all $z \in \mathbb{C}$. In particular it should be zero on the imaginary axis. On substituting for $p(y|z)$ and simplifying, we get
\begin{eqnarray}
-\gamma(\sigma^4-y^2)+(C-\gamma P) = \sum_{i=1}^{n}p_i \log{p(y+\mu_i)}
\end{eqnarray} 
which is not satisfied by density function of any random variable with infinite continuous support and $\gamma \geq 0$. Hence, it follows that the optimum distribution must have at most countable number of mass points. $\blacksquare$

%\newline

%\bibliographystyle{abbrv}
\bibliographystyle{IEEEtran}
\bibliography{mybibfilefade_modi1}

%\bibliography{IEEEabrv,mybibfile}
\end{document}